\documentclass{webofc}

\usepackage[varg]{txfonts}   
\usepackage{hyperref}
\usepackage{multirow}
\usepackage{url}
\usepackage{siunitx}
\hypersetup{colorlinks=true,citecolor=blue,urlcolor=blue,linkcolor=blue}

\DeclareSIUnit\clight{\text{\ensuremath{c}}}

\begin{document}
\title{traccc: GPU track reconstruction library for HEP experiments}
%
%

\author{\firstname{Paul}  \lastname{Gessinger} \inst{1} \and
        \firstname{Heather M.}  \lastname{Gray}\inst{2,3} \and
        \firstname{Attila}   \lastname{Krasznahorkay}\inst{4} \and
        \firstname{Charles}  \lastname{Leggett}\inst{3}  \and
        \firstname{Joana}    \lastname{Niermann}\inst{1} \and
        \firstname{Andreas}  \lastname{Salzburger}\inst{1}  \and
        \firstname{Stephen Nicholas} \lastname{Swatman}\inst{1}  \and
        \firstname{Beomki}   \lastname{Yeo}\inst{2,3}\fnsep\thanks{\email{beomki.yeo@berkeley.edu}}
}

\institute{European Organization for Nuclear Research, Meyrin, 1211, Switzerland
\and
Department of Physics, University of California, Berkeley, CA 94720, USA
\and
           Physics Division, Lawrence Berkeley National Laboratory, Berkeley, CA 94720, USA
\and
           Department of Physics, University of Massachusetts at Amherst, Amherst, MA 01003, USA 
          }

\abstract{We present the current development status and progress of traccc, a GPU track reconstruction library developed in the context of the \emph{A Common Tracking Software} (ACTS) project. traccc implements tracking algorithms used in high energy physics (HEP) experiments, including the Kalman filter--based track finding and fitting. 
We benchmark the software with data simulated by \textsc{Geant4} to measure the physics and computing performance. We show that the physics performance for GPU and CPU are very close. We also show that the GPUs can achieve higher computational performance than the CPU for sufficiently large events.
}

%
\maketitle
\section{Introduction}
\label{sec:introduction}

In collider experiments, charged particles created in particle collisions propagate through a detector with trajectories curved by a magnetic field. Their interactions with sensors allow their positions with respect to the location of the sensors to be measured. Track reconstruction refers to the process of computing the continuous trajectories of these particles from discrete measurements. Indeed, the track reconstruction is crucial in achieving high experimental sensitivity, which usually makes it the most computationally intensive stage of event reconstruction. 

Among microprocessors, the central processing unit (CPU) has long been the workhorse in the offline data analysis of most HEP experiments. The number of particles created per collision will be unprecedentedly high in future experiments such as those on the High Luminosity LHC (HL-LHC), and models of computation employing only CPUs are unlikely to meet the demands of such experiments~\cite{ATLAS2022_Roadmap}. Following the decline of Dennard scaling~\cite{DENNARD1974_DennardScaling} at the beginning of the twenty-first century, strategies for increasing the performance of microprocessors have shifted drastically from increasing clock frequencies---now impossible due to increased leakage currents---to increasing the number of cores, leading to the advent of parallel and massively parallel microprocessor architectures.

General purpose graphics processing units (GPUs) exemplify the shift towards parallel architectures, with many such devices featuring more than \num{10000} cores\footnote{GPU cores are not directly comparable to CPU cores in terms of performance, with a single CPU core delivering significantly more performance than a single GPU core.}. The massive parallelism of GPUs presents a promising way to meet the computational demands of future HEP experiments, but programming this many cores is challenging and algorithms that perform well on CPU-like architectures do not necessarily perform well on GPUs. We present traccc, a C++-based track reconstruction library for GPUs~\cite{GITHUB2025_traccc} and a part of the \emph{A Common Tracking Software} (ACTS) project \cite{AI2022_ACTS}. Even though the current development aims for the application to the ATLAS experiment of HL-LHC, traccc is designed to work on any type of detectors used in high-energy physics experiments. The algorithms and data layouts are modified in a GPU-friendly manner while traccc keeps the same physics performance of the original CPU codes.

Previous works \cite{GESSINGER2022_traccc, KRASZNAHORKAY2023_traccc} successfully demonstrated the early stages of the track reconstruction chain. However, other algorithms, e.g. track fitting based on Kalman Filter \cite{KALMAN1960_Kalman, FRUHWIRTH1987_Kalman}, remained incomplete due to the premature status in tracking geometry where tracks navigate through detector components efficiently. This work extends the scope of traccc to a full track reconstruction chain as the tracking geometry finally can be operated in GPU devices with precise track propagation algorithms. Its physics and computing performance are also investigated using the data simulated by Geant4 \cite{AGOSTINELLI2003_Geant4} in the Open Data Detector (ODD) \cite{GESSINGER2023_ODDACAT}.

\section{Implementation}
\label{sec:track_reconstruction_implementation}

The track reconstruction chain begins with the hit clustering where the positions of particles intersecting silicon-based sensors and their errors are calculated by finding the weighted centroids of clusters of activated cells. Clustering in track reconstruction is a connected component analysis (CCA) problem, but differs from most similar problems in the low density of its input, which is commonly less than \SI{1}{\percent}. In order to solve such sparse problems efficiently, we reduce to a CCA problem which we solve using the \textsc{FastSV}~\cite{ZHANG2020_FASTSV} algorithm. The set of a position and its error is called measurement, and its local position on the sensor is converted to a position in the global coordinate to form a spacepoint. The spacepoints are used in the seeding algorithm where triplets, a set of three spacepoints which can approximately form a helix, are found. In the track parameter estimation algorithm, the position and momentum of tracks, called a seed, are estimated at the first spacepoint using the helix model. 

The seeds are used as the initial track parameters for track propagation by a combinatorial Kalman filter (CKF) \cite{BILLOIR1989_CKF, BILLOIR1990_CKF2} where the full patterns of tracks are found. In the CKF, track parameters and their covariances are propagated based on the fourth order Runge-Kutta-Nystr\"{o}m method \cite{NYSTROM1925_RKN, BUGGE1981_RK, LUND2009_RKNPropagation, LUND2009_CovarianceTransport, YEO2024_JACOBIAN} and the Kalman Filter is applied to measurements on intersected sensor modules to calculate a $\chi^2$ value based on the track parameter covariances and spatial residual between the track and measurement. If the $\chi^2$ is small enough, the measurement is attached to the track to make a new branch for the propagation to the next sensor module. The track parameters and their covariances updated by Kalman Filter during the CKF are most accurate at the last measurement as Kalman Filter accumulates all information of previous measurements. However, the filtered track parameters at the first measurement, which is important for vertex reconstruction, are estimated only with one measurement hence least accurate. Therefore, a Kalman Filter-based fitting \cite{FRUHWIRTH1987_Kalman}, which runs a smoothing algorithm called the Rauch--Tung--Striebel (RTS) smoother \cite{RAUCH1965_RTS}, is applied additionally to the tracks to make all track parameters consider all measurements in the pattern.

The algorithms are developed for various GPU computing platforms, mainly in CUDA and SYCL \cite{REYES2015_SYCL}, as well as on CPU for comparison purposes. CUDA, the most widely used GPU computing platform, is exclusive to NVIDIA devices while SYCL codes can also be operated on AMD and Intel devices. Floating point arithmetic is implemented according to the IEEE~754 standard in both single- and double-precision, which we will henceforth refer to as FP32 and FP64, respectively. Although FP64 has been a common choice in HEP computing due to the increased precision it provides, FP32 is investigated as well because GPUs are much more efficient with FP32.


The track reconstruction chain is designed to minimize data transfers between host and device. For every event, the data transfers occur when the pixel hit data are transferred to GPU and when fitted track parameters are retrieved from GPU. The transfer of tracking geometry and magnetic field to GPU needs to happen only once at the first event. Multiple CPU threads can be used to operate the multiple GPU pipelines asynchronously, where each of CPU thread handles one invocation of the GPU track reconstruction. This enables reconstructing multiple events concurrently as a single GPU pipeline running on a single event may not be enough to utilize all resources of GPU device.

\section{Performance}
\label{sec:performance}

To evaluate physics and computing performance, data are prepared by simulating $t\bar{t}$ collision events in the ODD. The ODD, whose design is inspired by the inner tracker of the ATLAS experiment at the HL-LHC \cite{GONELLA2023_ITk}, consists of---in order of increasing distance from the interaction point---a pixel detector, a short-length strip detector, and a long-length strip detector. Along the longitudinal direction, detector modules are arranged in the barrel detector at the center and in the endcap detector at both ends. A uniform magnetic field of \SI{2}{\tesla} is applied along the longitudinal ($z$) axis. During reconstruction, the effects from material interactions are evaluated by assuming muon trajectories. 

In the following subsections, the GPU computations are validated against the CPU one, and the efficiency of track reconstruction is studied. The computing performance of the GPU relative to the CPU is also presented. All GPU results are obtained using NVIDIA GPUs with the CUDA implementation.

\begin{table}
\centering
\caption{Comparison of CPU and GPU computing results for FP32 and FP64. Track patterns are considered matched if all measurements are identical. \textit{Tolerated Diff.} denotes the maximum relative difference aloowed for track parameters to match.}
\label{tab:validation}      
\begin{tabular}{c|ccc}
\hline
   & Tolerated Diff. & Matching ratio (FP32) & Matching ratio (FP64) \\
   \hline
CKF pattern   & - & 94.7\% &  99.4\% \\
   \hline
       & 0.01\%  & 37.8\% & 91.8\% \\
Fitted    & 0.1\%   & 52.4\% & 93.5\% \\
   parameters    & 1\%     & 65.9\% & 95.3\% \\
       & 5\%     & 75.4\% & 96.5\% \\
   \hline
\end{tabular}
\end{table}

\subsection{Computing validation}
\label{subsec:comp_validation}

The results of GPU computation can differ from those of CPU computations for the following reasons: GPUs execute multiplication and addition operations concurrently, whereas CPUs typically perform them separately. In addition, GPU computations may exhibit non-deterministic behavior due to atomic operations, where multiple threads read from and write to the same variable in an arbitrary order, potentially resulting in varying rounding errors. Table \ref{tab:validation} presents the matching ratios between GPU and CPU results, based on comparisons of CKF track finding patterns and fitted parameters. The track finding patterns show good agreement for both FP32 and FP64, even though the matching condition requires exact pattern identity. The non-negligible discrepancies observed in the fitted parameters for FP32 may be attributed to the known numerical instability of the RTS smoother \cite{FRUHWIRTH2000_DataAnalysis, FRUHWIRTH2021_PatternRecognition}.

\begin{figure*}
\centering
\vspace*{1cm}       
\includegraphics[width=6cm,clip]{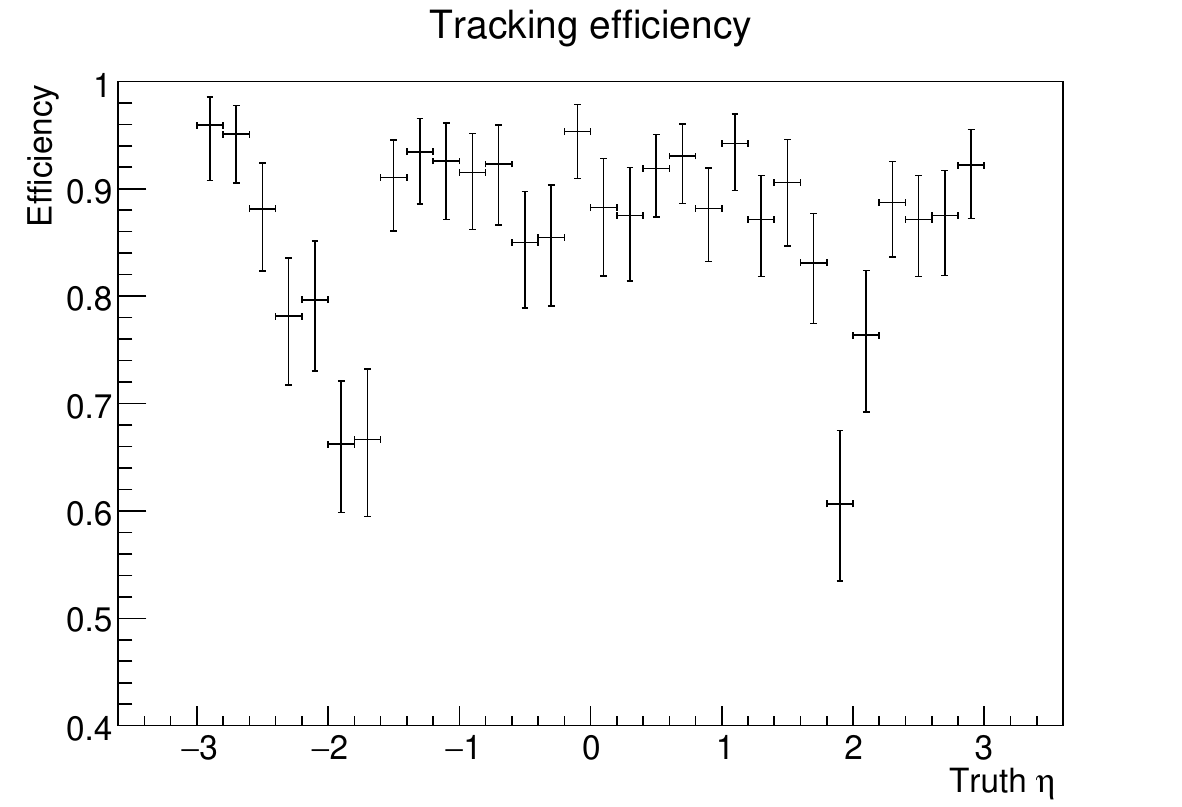}
\caption{Tracking efficiency of the CPU implementation as a function of $\eta$ for a $t\bar{t}$ collision event in the ODD geometry. The event is simulated with 140 pileup and the FP32 precision is used.}
\label{fig:efficiency}      
\end{figure*}

Tracking efficiency, defined as the fraction of successfully reconstructed particles among reconstructible ones, is also evaluated to validate the performance of CKF. A particle is considered reconstructible if it carries electric charge, has a transverse momentum greater than \SI[per-mode=symbol]{500}{\mega\eV\per\clight}, and originates from a vertex located within the barrel region of the pixel detector corresponding to $|z|<$ \SI{500}{\mm} and $r<$ \SI{200}{\mm}. To be considered successfully reconstructed, a track must satisfy the so-called \textit{double matching condition} with respect to any reconstructible particles. The \textit{major} particle of a found track is the one that contributes the largest number of measurements to the track pattern. We define $N$ as the number of measurements from the major particle included in the track, $N_t$ as the total number of measurements in the found track, and $N_p$ as the total number of measurements produced by the major particle. The double matching condition requires that $N > N_t/2$ and $N > N_p/2$. Fig. \ref{fig:efficiency} shows the tracking efficiency as a function of pseudorapidity ($\eta$) of particles in $t\bar{t}$ collision events with 140 pileup, computed using FP32. It should be noted that the efficiency is measured using the CPU implementation, thus, the efficiency with the GPU implementation may differ up to 5\% according to Table \ref{tab:validation}.  The dips at $|\eta| \approx 2$ arise from particles traversing the transition region between the barrel and endcap detectors, where track propagation is more likely to miss detector modules.

\begin{table}
\centering
\caption{Specifications of hardware devices used in computing performance benchmark. \textit{FP64:FP32} indicates the theoretical performance ratio between FP64 and FP32 operations. \textit{Memory} refers to the size of the DRAM available on each device.}
\label{tab:specs}       
\begin{tabular}{c|cccccc}
\hline
   & Model          & Cores & Memory & FP32 FLOPS & FP64:FP32  \\
   \hline
       & NVIDIA RTX A6000      & \num{10752} & 48 GB  & 38.7 TFLOPS & 1:32\\
 GPU   & NVIDIA RTX 2080 SUPER & \num{3072}  & 8 GB   & 11.2 TFLOPS & 1:32 \\
       & NVIDIA A30            & \num{3584}  & 24 GB  & 10.3 TFLOPS & 1:2 \\
   \hline
CPU   & AMD EPYC 7413  & \num{24}    &  -     & 1.02 TFLOPS & 1:2 \\
   \hline
\end{tabular}
\end{table}

\subsection{Computing performance}
\label{subsec:comp_performance}

For the computing performance benchmark, GPU models were selected to cover a wide range of hardware capabilities. Table \ref{tab:specs} summarized the specifications of the devices used in the benchmark. The RTX A6000 delivers the highest FP32 performance but shows a significant drop in FP64 performance. The RTX 2080 SUPER, a consumer-grade graphics card, offers a balanced trade-off between price and FP32 performance. The A30 is distinct from the other two GPUs in that its FP64 performance is half of its FP32 performance. Although all GPU models outperform the CPU in terms of raw FP32 FLOPS, it is important to note that FLOPS do not necessarily reflect real-world performance. A CPU core is generally more capable than a GPU core for complex computations, owing to its larger cache size and features such as branch prediction. In contrast, GPU threads execute in groups called \textit{warp}, which operate in lock-step meaning that all threads in a warp must execute the same instruction simultaneously. As a result, GPU performance can suffer in the presence of divergent control logic.

\begin{figure}[ht]
\centering
\includegraphics[width=6cm,clip]{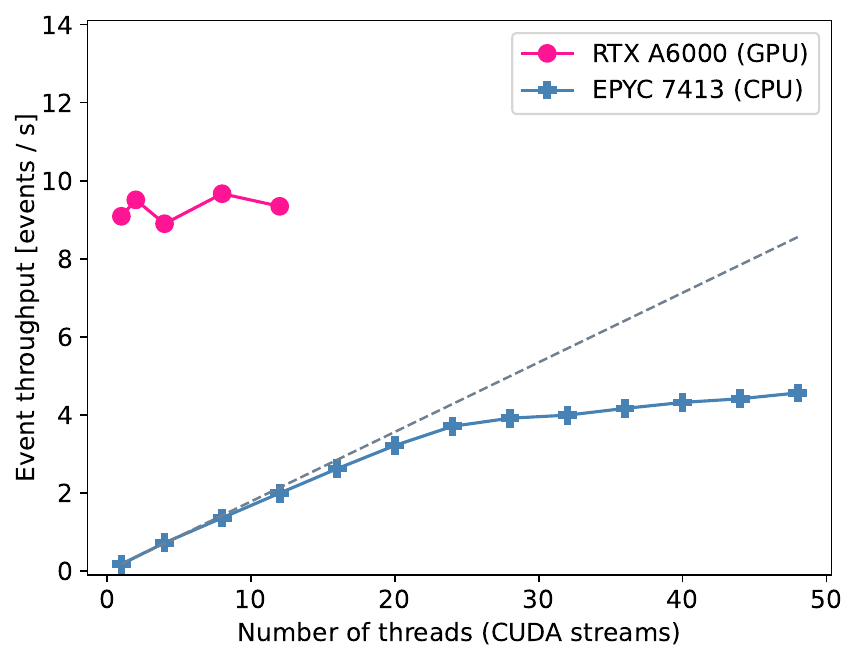}
\caption{Event throughput measured with the RTX A6000 (GPU) and AMD EPYC 7413 (CPU) as a function of the number of CPU threads. $t\bar{t}$ collision events with 140 pileup are reconstructed using FP32 precision. The dotted line represents ideal linear scaling of CPU throughput with increasing thread count.}
\label{fig:throughput_vs_threads}   
\end{figure}

As mentioned in Section \ref{sec:track_reconstruction_implementation}, the computing speed of both GPU and CPU implementations can be improved by reconstructing multiple events concurrently using multiple CPU threads. In the CUDA implementation, each CPU thread launches a separate CUDA \textit{stream} which handles the reconstruction pipeline for a single event. Fig. \ref{fig:throughput_vs_threads} shows the event throughput---the number of events reconstructed per unit time---for the RTX A6000 and AMD EPYC 7413, as a function of the number of CPU threads. The CPU throughput scales nearly linearly with the number of logical cores (i.e., up to 24), with modest gains beyond that due to simultaneous multithreading, where each core handles two threads. On the contrary, the GPU shows no clear improvement in throughput with additional concurrent events, likely due to inefficiencies in GPU memory allocation.

\begin{figure*}
\centering
\vspace*{1cm}       
\includegraphics[width=5.5cm,clip]{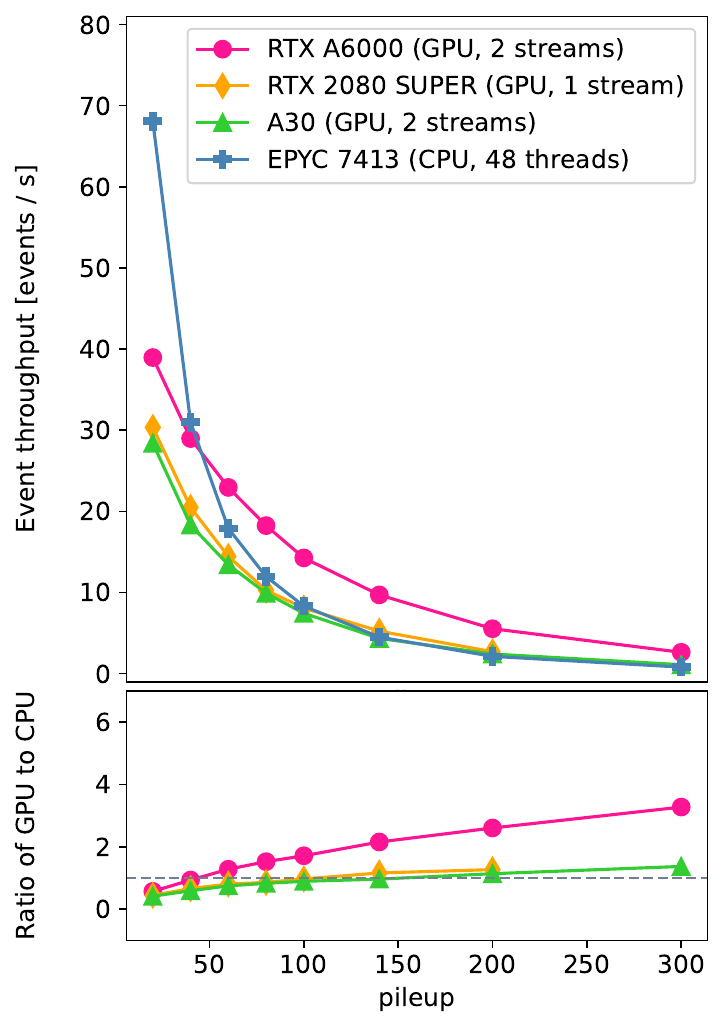}
\includegraphics[width=5.5cm,clip]{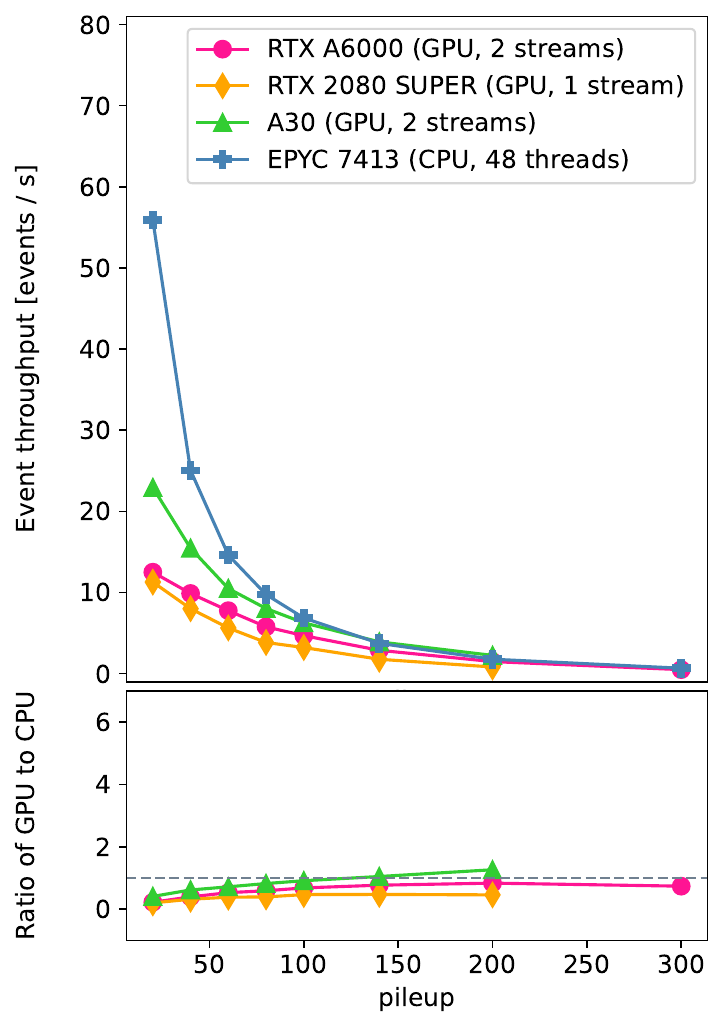}
\caption{Event throughput as a function of $t\bar{t}$ pileup for FP32 (left) and FP64 (right) precision. The bottom plots show the ratio of GPU to CPU event throughput.}
\label{fig:throughput_vs_pileups}    
\end{figure*}

The event throughput of GPUs and CPUs as a function of $t\bar{t}$ pileup is presented in Fig. \ref{fig:throughput_vs_pileups}. Based on the results of Fig. \ref{fig:throughput_vs_threads}, the number of CPU threads, or CUDA streams, is fixed at 48 and 2 for CPU and GPU pipelines, respectively. An exception is made for the RTX 2080 SUPER, which uses only one CUDA stream due to its limited memory capacity of 8 GB. In the FP32 benchmark, the RTX A6000 outperforms the CPU in most pileup scenarios, while the other GPUs only exceed CPU performance when the pileup exceeds 200. Overall, the relative performance between GPU and CPU is smaller than the FP32 FLOPS ratio reported in Table \ref{tab:specs}. This discrepancy results from the aforementioned advantages of CPU cores and the logic divergence in the CKF algorithm, where track propagation is not evenly balanced across threads. For the FP64 benchmark, only A30, which has the highest FP64-to-FP32 performance ratio in Table \ref{tab:specs}, surpasses the CPU performance only under high pileup conditions.

\section{Discussion}
\label{sec:discussion}

While the current track finding algorithm demonstrates good tracking efficiency, a complete understanding of its physical performance requires investigation of duplicate and fake track rates. Duplicate tracks arise when multiple tracks share the same or very similar hit patterns, whereas fake tracks correspond to patterns that do not match any truth particle. To address these issues, an additional algorithm---known as ambiguity resolution---will be integrated into the reconstruction chain to remove such tracks after the CKF stage.

Although the cause of the numerical instability of the RTS smoother in FP32 remains unclear, alternative smoothing algorithms may offer improved matching rates in Kalman Filter-based fitting. One such candidate is the \textit{optimal smoother} \cite{FRASER1967_OptimalSmoother, FRUHWIRTH2021_PatternRecognition}, which computes the smoothed track parameters as a weighted average of the parameters from both forward and backward filters. This method, known to be more stable, is currently under development for future integration.

There remains room for optimization in GPU computing performance, and several improvements are already underway. The weak scaling behavior with respect to multiple CUDA streams can be addressed by enhancing memory allocation efficiency. As an alternative approach, multiple GPU pipelines can be deployed using an \textit{as-a-service} model, such as Triton \cite{GITHUB2025_Triton} server of NVIDIA. The primary performance bottleneck from the logic divergence in the CKF track finding is also being targeted for improvement. By restructuring the track propagation steps and distributing them more evenly accross threads within a warp, it is expected that event throughput can be increased by several factors~\cite{SWATMAN2022_LossEvaluation}.

\section{Summary and conclusions}
\label{sec:summary}

GPUs are expected to accelerate HEP data analysis, which has been limited by the slow performance growth of traditional CPUs. Among various processing steps, track reconstruction is the most computationally intensive and inherently parallelizable, making it an ideal candidate for GPU acceleration. To address this need, a general GPU tracking library, traccc, has been developed.

In this study, both CPU and GPU-based track reconstruction chains were tested using simulated data in the ODD geometry. The CKF algorithm demonstrated good tracking efficiency; however, further studies are needed to quantify duplicate and fake tracks, and to incorporate an ambiguity resolution algorithm to mitigate them. The low matching rate between CPU and GPU in the fitted track parameters also indicates the need for replacing current smoother with a more stable alternative. In terms of the computing performance, GPUs outperform CPUs in high pileup conditions when using the FP32 precision. Significant further improvements are anticipated through more efficient concurrent event reconstruction using multiple GPU pipelines, and through optimization of track propagation in the CKF algorithm.

\section*{Acknowledgements}
This work was supported by the National Science Foundation under Cooperative Agreements (OAC-1836650, PHY-2323298 and PHY-2120747), the CERN Strategic R\&D Program on Technologies for Future Experiments (CERN-OPEN-2018-006), the Eric \& Wendy Schmidt Fund for Strategic Innovation through the CERN Next Generation Triggers project (grant agreement number SIF-2023-004) and the Wolfgang Gentner Programme of the German Federal Ministry of Education and Research (grant no. 13E18CHA). Proofreading assistance was provided by ChatGPT.

\bibliography{references}

\begin{thebibliography}{26}

\bibitem{ATLAS2022_Roadmap}
{ATLAS Collaboration}, {ATLAS Software and Computing HL-LHC Roadmap} (2022), \urlstyle{tt}\url{https://cds.cern.ch/record/2802918}

\bibitem{DENNARD1974_DennardScaling}
R.~Dennard et~al., {Design of ion-implanted MOSFET's with very small physical dimensions}, IEEE Journal of Solid-State Circuits \textbf{9}, 256 (1974). \doiwoc{https://doi.org/10.1109/JSSC.1974.1050511}

\bibitem{GITHUB2025_traccc}
B.~Yeo et~al., Github repository of acts-project/traccc, \urlstyle{tt}\url{https://github.com/acts-project/traccc}

\bibitem{AI2022_ACTS}
X.~Ai et~al., {A Common Tracking Software Project}, Comput. Softw. Big Sci. \textbf{6}, 8 (2022). \doiwoc{https://doi.org/10.1007/s41781-021-00078-8}

\bibitem{GESSINGER2022_traccc}
P.~Gessinger et~al., {ACTS GPU Track Reconstruction Demonstrator for HEP}, Proceedings of the CTD 2022 pp. {46--53} (2023). \doiwoc{https://doi.org/10.5281/zenodo.8119864}

\bibitem{KRASZNAHORKAY2023_traccc}
A.~Krasznahorkay et~al., {traccc - a close to single-source track reconstruction demonstrator for CPU and GPU}, \url{https://indico.jlab.org/event/459/contributions/11420/}, presented at the 26th International Conference on Computing in High Energy and Nuclear Physics

\bibitem{KALMAN1960_Kalman}
R.E. Kalman, {A New Approach to Linear Filtering and Prediction Problems}, Journal of Basic Engineering \textbf{82}, 35 (1960). \doiwoc{https://doi.org/10.1115/1.3662552}

\bibitem{FRUHWIRTH1987_Kalman}
R.~Frühwirth, {Application of Kalman filtering to track and vertex fitting}, Nucl. Instr. and Meth. A \textbf{262}, 444 (1987). \doiwoc{https://doi.org/10.1016/0168-9002(87)90887-4}

\bibitem{AGOSTINELLI2003_Geant4}
S.~Agostinelli et~al., Geant4—a simulation toolkit, Nucl. Instr. and Meth. A \textbf{506}, 250 (2003). \doiwoc{https://doi.org/10.1016/S0168-9002(03)01368-8}

\bibitem{GESSINGER2023_ODDACAT}
P.~Gessinger-Befurt, A.~Salzburger, J.~Niermann, {The Open Data Detector Tracking System}, Journal of Physics: Conference Series \textbf{2438}, 012110 (2023). \doiwoc{https://doi.org/10.1088/1742-6596/2438/1/012110}

\bibitem{ZHANG2020_FASTSV}
Y.~Zhang, A.~Azad, Z.~Hu, FastSV: A Distributed-Memory Connected Component Algorithm with Fast Convergence (2020), pp. 46--57

\bibitem{BILLOIR1989_CKF}
P.~Billoir, {Progressive track recognition with a Kalman-like fitting procedure}, Computer Physics Communications \textbf{57}, 390 (1989). \doiwoc{https://doi.org/10.1016/0010-4655(89)90249-X}

\bibitem{BILLOIR1990_CKF2}
P.~Billoir, S.~Qian, {Simultaneous pattern recognition and track fitting by the Kalman filtering method}, Nucl. Instr. and Meth. A \textbf{294}, 219 (1990). \doiwoc{https://doi.org/10.1016/0168-9002(90)91835-Y}

\bibitem{NYSTROM1925_RKN}
E.~Nystr{\"o}m, {{\"U}ber die numerische Integration von Differentialgleichungen}, Acta Societatis scientiarum Fennicae (Druck der Finnischen literaturgesellschaft, 1925)

\bibitem{BUGGE1981_RK}
L.~Bugge, J.~Myrheim, {Tracking and track fitting}, Nuclear Instruments and Methods \textbf{179}, 365 (1981). \doiwoc{https://doi.org/10.1016/0029-554X(81)90063-X}

\bibitem{LUND2009_RKNPropagation}
E.~Lund, L.~Bugge, I.~Gavrilenko, A.~Strandlie, {Track parameter propagation through the application of a new adaptive Runge-Kutta-Nyström method in the ATLAS experiment}, JINST \textbf{4}, P04001 (2009). \doiwoc{https://doi.org/10.1088/1748-0221/4/04/P04001}

\bibitem{LUND2009_CovarianceTransport}
E.~Lund, L.~Bugge, I.~Gavrilenko, A.~Strandlie, {Transport of covariance matrices in the inhomogeneous magnetic field of the ATLAS experiment by the application of a semi-analytical method}, JINST \textbf{4}, P04016 (2009). \doiwoc{https://doi.org/10.1088/1748-0221/4/04/P04016}

\bibitem{YEO2024_JACOBIAN}
B.~Yeo, H.~Gray, A.~Salzburger, S.N. Swatman, {The derivation of Jacobian matrices for the propagation of track parameter uncertainties in the presence of magnetic fields and detector material}, Nucl. Instr. and Meth. A \textbf{1068}, 169734 (2024). \doiwoc{https://doi.org/10.1016/j.nima.2024.169734}

\bibitem{RAUCH1965_RTS}
H.E. Rauch, F.~Tung, C.T. Striebel, {Maximum likelihood estimates of linear dynamic systems}, AIAA Journal \textbf{3}, 1445 (1965). \doiwoc{https://doi.org/10.2514/3.3166}

\bibitem{REYES2015_SYCL}
R.~Reyes, V.~Lom{\"u}ller, SYCL: Single-source C++ accelerator programming, in \emph{International Conference on Parallel Computing} (2015)

\bibitem{GONELLA2023_ITk}
L.~Gonella, {The ATLAS ITk detector system for the Phase-II LHC upgrade}, Nucl. Instr. and Meth. A \textbf{1045}, 167597 (2023). \doiwoc{https://doi.org/10.1016/j.nima.2022.167597}

\bibitem{FRUHWIRTH2000_DataAnalysis}
R.~Frühwirth et~al., {Data analysis techniques for high-energy physics; 2nd ed.}, Cambridge monographs on particle physics, nuclear physics, and cosmology (Cambridge Univ. Press, Cambridge, 2000)

\bibitem{FRUHWIRTH2021_PatternRecognition}
R.~Frühwirth, A.~Strandlie, {Pattern Recognition, Tracking and Vertex Reconstruction in Particle Detectors} (Springer Cham, 2021)

\bibitem{FRASER1967_OptimalSmoother}
D.C. Fraser, Ph.D. thesis, Massachusetts Institute of Technology (1967), \urlstyle{tt}\url{http://hdl.handle.net/1721.1/13543}

\bibitem{GITHUB2025_Triton}
{\relax NVIDIA Corporation}, Triton inference server: An optimized cloud and edge inferencing solution, \urlstyle{tt}\url{https://github.com/triton-inference-server}

\bibitem{SWATMAN2022_LossEvaluation}
S.N. Swatman, A.L. Varbanescu, A.~Krasznahorkay, A.~Pimentel, Modelling Performance Loss due to Thread Imbalance in Stochastic Variable-Length SIMT Workloads, in \emph{2022 30th International Symposium on Modeling, Analysis, and Simulation of Computer and Telecommunication Systems (MASCOTS)} (2022), pp. 137--144

\end{thebibliography}

\end{document}